\documentclass[12pt, prd, showpacs]{revtex4}
\usepackage{amssymb}
\usepackage{amsmath}

\input{tcilatex}

\begin{document}

\title{Acceleration of particles by rotating black holes: near-horizon
geometry and kinematics}
\author{O. B. Zaslavskii}
\affiliation{Department of Physics and Technology, Kharkov V.N. Karazin National
University, 4 Svoboda Square, Kharkov, 61077, Ukraine}
\email{zaslav@ukr.net }

\begin{abstract}
Nowadays, the effect of infinite energy in the centre of mass frame due to
near-horizon collisions attracts much attention.We show generality of the
effect combining two seemingly completely different approaches based on
properties of a particle with respect to its local light cone and
calculating its velocity in the locally nonrotaing frame directly. In doing
so, we do not assume that particles move along geodesics. Usually, a
particle reaches a horizon having the velocity equals that of light.
However, there is also case of "critical" particles for which this is not
so. It is just the pair of usual and critical particles that leads to the
effect under discussion. The similar analysis is carried out for massless
particles. Then, critical particles are distinguishable due to the
finiteness of local frequency. Thus, both approach based on geometrical and
kinematic properties of particles moving near the horizon, reveal the
universal character of the effect.
\end{abstract}

\maketitle

\section{Introduction}

The effect of infinite grow of the energy in the centre of mass frame due to
the near-horizon collision of two particles (BSW effect) attracts now much
attention \cite{ban} - \cite{op}. At first, it was discovered for the Kerr
metric \cite{ban} but later it was understood that the effect is of quite
general character. It was shown in \cite{cqg} that such an effect exists if
orientation of a four-velocity of a massive particle with respect to its
local light cone obeys some simple conditions. More precisely, one of
coefficients in a suitable null tetrad basis (which is specified below)
should vanish on the horizon ("critical" particle). From the other hand, in
recent works \cite{k}, an alternative explanation was given in terms of
kinematic properties of particles. It turned out that from a kinematic
viewpoint, a critical particle is distinguished by the property that in the
horizon limit, its velocity in the locally nonrotating frame (LNRF) \cite%
{aj72} tends to the value which is less than that of light. Meanwhile, for
typical ("usual") particles this velocity tends just to the speed of light.
It is collision between a critical and usual particles that produces the
effect under discussion if such a collision occurs near the horizon.

Qualitatively, it can be explained as follows. If we have two particles one
of which is slow (the value of the velocity $v_{1}<1$, the speed of light $%
c=1$) and the second one is fast ($v_{2}\approx 1$), the relative velocity
is also close to $1$. Correspondingly, the Lorentz factor tends to infinity
and we have the BSW effect. Actually, this is explained in terms of simple
kinematics of collision in the flat space-time. Further use of the kinematic
approach for investigation of rather subtle details of the BSW effect in the
Kerr background can be found in Ref. \cite{op}. If one of particles is
massless, so its velocity is equal to $1$ exactly, the explanation is
somewhat changed. It is based on the relative role of the gravitational
blueshift and the Doppler effect. In doing so, the usual and critical
particles are distinguished by the property that the LNRF frequency of the
massless particle is finite or infinite . It turned out that collision
between massive and massless particles produce the BSW effect also for the
situation when one particle is critical and the other one is usual \cite{k}.

Thus, we have two quite different explanations using different language -
from the geometric point of view and on the basis of kinematics. The
geometric explanation \cite{cqg} was quite general whereas the kinematic one 
\cite{k}, was obtained for geodesic motion of particles only.

The aim of the present work is (i) to make a bridge between geometric and
kinematic approaches and (ii) generalize kinematic one to an arbitrary case
not requiring geodesic motion and not using equations of particles' motion
at all.

\section{Basic equations}

Let us consider the space-time of a rotating black hole described by the
metric

\begin{equation}
ds^{2}=-N^{2}dt^{2}+g_{\phi \phi }(d\phi -\omega dt)^{2}+dl^{2}+g_{zz}dz^{2}.
\label{m}
\end{equation}%
The location of the horizon defined as a surface of inifnite redshift
corresponds to $N\rightarrow 0$. In what follows we will use the tetrad
basis. Denoting coordinates $x^{\mu }$ as \ $x^{0}=t,x^{1}=l$, $x^{2}=z$, $%
x^{3}=\phi $, we choose the orthonormal tetrad vectors $h_{(a)\mu }$ in the
following way:%
\begin{equation}
h_{(0)\mu }=-N(1,0,0,0)\text{, }  \label{h0}
\end{equation}%
\begin{equation}
h_{(1)\mu }=(0,1,0,0)
\end{equation}%
\begin{equation}
h_{(2)\mu }=\sqrt{g_{zz}}(0,0,1,0)
\end{equation}%
\begin{equation}
h_{(3)\mu }=\sqrt{g_{\phi \phi }}(-\omega ,0,0,1)  \label{h3}
\end{equation}%
If such a tetrad is attached to an observer moving in the metric (\ref{m}),
it has meaning of zero angular momentum observer \cite{aj72}, so two
abbreviations LNRF and ZAMO are used in literature. A corresponding observer
"rotates with the geometry" in the sense that $\frac{d\phi }{dt}\equiv
\omega $ for him. The advantage of using the tetrad components consists in
that one can use the formulas of special relativity in the flat space-time
tangent to any given point.

In a given context \cite{cqg}, a null tetrad is also convenient for the
decomposition of the metric: 
\begin{equation}
g_{\alpha \beta }=-l_{\alpha }N_{\beta }-l_{\beta }N_{\alpha }+\sigma
_{\alpha \beta }  \label{nt}
\end{equation}%
where $\sigma _{\alpha \beta }=a_{\alpha }a_{\beta }+b_{\alpha }b_{\beta }$, 
$l^{\alpha }\sigma _{\alpha \beta }=N^{\alpha }\sigma _{\alpha \beta }=0$, $%
a_{\mu }$ and $b_{\mu }$ are spacelike vectors (see, for example, textbook 
\cite{erik}). \ For the metric (\ref{m}) one can check that the null vectors
can be chosen in the following way:%
\begin{equation}
l_{\mu }=(-N^{2},N,0,0),  \label{7}
\end{equation}%
\begin{equation}
N_{\mu }=\frac{1}{2}(-1,-\frac{1}{N},0,0).  \label{8}
\end{equation}%
\begin{equation}
(Nl)=-1\text{.}
\end{equation}

Then, it is seen that%
\begin{equation}
h_{(0)\mu }=NN_{\mu }+\frac{l_{\mu }}{2N},
\end{equation}%
\begin{equation}
h_{(1)\mu }=\frac{l_{\mu }}{2N}-NN_{\mu }.
\end{equation}

\section{Case of massive particles}

Let us consider motion of massive particles (electrons, for brevity). Then,
using our tetrad basis, we can write down the decomposition of the
four-velocity,%
\begin{equation}
u^{\mu }=\frac{l^{\mu }}{2\alpha }+\beta N^{\mu }+s^{\mu }\text{, }s^{\mu
}=Aa^{\mu }+Bb^{\mu }\text{,}  \label{u}
\end{equation}%
$A$ and $B$ are coefficients. The normalization condition $(uu)=-1$ entails%
\begin{equation}
\alpha =\frac{\beta }{(ss)+1}.  \label{cu}
\end{equation}%
The vector $u^{\mu }$ is future-directed. We consider the vicinity of the
future horizon, so vectors $l^{\mu }$ and $N^{\mu }$ are also
future-directed. Therefore, in what follows the coefficients $\alpha \geq 0$%
, $\beta \geq 0$.

Straightforward calculations give us%
\begin{equation}
-(uh_{(0)})=\frac{\alpha \beta +N^{2}}{2N\alpha },
\end{equation}%
\begin{equation}
(uh_{(1)})=\frac{N^{2}-\alpha \beta }{2\alpha N}.
\end{equation}%
\begin{equation}
(uh_{(2)})=(h_{(2)}s)\text{,}
\end{equation}

\begin{equation}
(uh_{(3)})=\frac{L}{\sqrt{g_{\phi \phi }}}
\end{equation}%
where $L=u_{\phi }$ is the angular momentum per unit mass. If the metric
does not depend on $\phi $, it is conserved. However, we do not exploit such
a property, so our consideration is more general.

Then, we can introduce the three-velocity in this frame according to \cite%
{aj72}:%
\begin{equation}
v^{(i)}=v_{(i)}=\frac{u^{\mu }h_{\mu (i)}}{-u^{\mu }h_{\mu (0)}}.  \label{vi}
\end{equation}%
The absolute value of the velocity equals%
\begin{equation}
v^{2}=\left[ v^{(1)}\right] ^{2}+\left[ v^{(2)}\right] ^{2}+\left[ v^{(3)}%
\right] ^{2}\text{.}  \label{v2}
\end{equation}

It is seen from (\ref{vi}) that $v^{2}<1$ as it should be for massive
particles. Indeed, using the representaion of the metric in terms of
orthonormal tetrad 
\begin{equation}
g_{\mu \nu }=-h_{(0)\mu }h_{(0)\nu }+h_{(i)\mu }h_{(i)\nu }
\end{equation}%
where summation is taken over index $i$ and taking into account that
\thinspace $(uu)=-1$, one obtains that%
\begin{equation}
v^{2}=\frac{\varepsilon }{1+\varepsilon }<1  \label{less}
\end{equation}%
where $\varepsilon =u^{\mu }u^{\nu }h_{(i)\mu }h_{\nu }^{(i)}>0$. Actually,
eq. (\ref{vi}) is nothing else than natural generaization of formulas of
special relativity $v^{i}=\frac{u^{i}}{u^{0}}$ where $u^{0}=\frac{1}{\sqrt{%
1-v^{2}}}$.

One can check that for our metric and choice of tetrads

\begin{equation}
v^{(1)}=\frac{N^{2}-\alpha \beta }{N^{2}+\alpha \beta }\text{,}  \label{v1}
\end{equation}%
\begin{equation}
v^{(2)}=2(h_{(2)}s)\frac{N\alpha }{\alpha \beta +N^{2}}\text{,}
\end{equation}%
\begin{equation}
v^{(3)}=\frac{2LN}{\sqrt{g_{\phi \phi }}}\frac{\alpha }{N^{2}+\alpha \beta }%
\text{.}  \label{v3}
\end{equation}%
In the particular case of the Kerr metric, the effect of infinite
acceleration for geodesic particle with all nonzero components of the
velocity was studied in \cite{kerr}.

In the horizon limit.$N\rightarrow 0$ we obtain for a generic case ("usual"
particles) that 
\begin{equation}
v^{(1)}\rightarrow -1\text{, }v^{(2)}\rightarrow 0\text{, }%
v^{(3)}\rightarrow 0\text{, }v\rightarrow 1\text{. }
\end{equation}%
Here the sign "minus" corresponds to motion towards the horizon.

However, there is special case ("critical particles") when near the horizon
the quantity $\beta \rightarrow 0$ when $N\rightarrow 0$. As for a
space-like vector $(ss)>0$, the denominator in (\ref{cu})\ does not vanish,
so $\alpha $ has the same order as $\beta $. We assume that the first
nonvanishing term in the Taylor expansion of $\beta $ has the order $N.$
(This is confirmed by explicit calculations for the geodesic motion in the
Kerr metric \cite{cqg}. In general, this can be taken simply as an
assumption that, by definition, distinguishes usual and critical particles.)
If $\beta \approx c_{1}N$, the coefficient $\alpha \approx c_{2}N$ where $%
c_{1}$ and $c_{2}$ are some coefficients. As, as is explained above, the
coefficients $\alpha $ and $\beta $ cannot be negative and $N>0$ by
definition, the coefficients $c_{1,2}>0$.

Then, it follows from (\ref{v1}), (\ref{v3}) that in the limit under
discussion%
\begin{equation}
\left\vert v^{(1)}\right\vert \rightarrow \frac{1-c_{1}c_{2}}{1+c_{1}c_{2}}<1%
\text{.}
\end{equation}%
\begin{equation}
v^{(2)}\rightarrow 2(sh_{(2)})\frac{c_{2}}{c_{1}c_{2}+1}\text{,}
\end{equation}%
\begin{equation}
v^{(3)}\rightarrow \frac{2L}{\sqrt{g_{\phi \phi }}}\frac{c_{2}}{1+c_{1}c_{2}}%
\text{.}
\end{equation}%
Thus, both components have the same order, \ the particle hits the horizon
nonperpendicularly (as was noticed in \cite{ban} for the Kerr metric), $%
v\neq 1$. Actually, this means that $v<1$ according to the property (\ref%
{less}).

Let us denote the absolute values of velocities as $v_{1}$ and $v_{2}$ for
particles 1 and 2, respectively (not to be confused with the tetrad
components). Once the properties $v_{1}\rightarrow 1$, $v_{2}<1$ are
establish for some pairs of particles, the further analysis of their
collisions which was elaborated in \cite{k} applies to this case directly,
so we reduce our problem to the known one. As a result, the relative
velocity of such two particles $w\rightarrow 1$, the corresponding Lorentz
factor diverges and we gain an infinite energy in the centre of mass frame
(see \cite{k} for details).

Some reservations are in order. Particles can approach the extremal horizon
but in the critical case cannot reach it. Then, the proper time needed for
this is infinite. If the horizon is nonextremal, the critical particle
cannot penetrate the potential barrier but the near-critical one can
approach the horizon as nearly as one like. Then, the energy of collision is
finite but can be made as large as one wishes. These issues are already
considered in \cite{gp4}, \cite{gpm}, \cite{prd}, so we do not repeat
details here.

\section{Massless case}

A\ massless particle (photon for brevity) is characterized by the wave
four-vector $k^{\mu }$. Then, the frequency measured by ZAMO equals%
\begin{equation}
\omega =-k^{\mu }h_{\mu (0)}\text{.}
\end{equation}

Now, the normalization condition changes to $(kk)=0$, so instead of (\ref{cu}%
) we have%
\begin{equation}
\alpha =\frac{\beta }{(ss)}\text{.}
\end{equation}

Then,%
\begin{equation}
\omega =\frac{\alpha \beta +N^{2}}{2N\alpha }
\end{equation}

For usual photons, with $\alpha ,\beta \neq 0$,%
\begin{equation}
\omega \approx \frac{\beta }{2N}\rightarrow \infty \text{.}
\end{equation}

For critical ones, again $\alpha \sim \beta $ near the horizon, $\beta
\approx c_{1}N$, the coefficient $\alpha \approx c_{2}N$, so we obtain that%
\begin{equation}
\omega \approx \frac{c_{1}c_{2}+1}{2c_{2}}
\end{equation}%
remains finite. This is nothing else than, by definition, the critical
photons.

Once the existence of such photnos with finite $\omega $ near the horizon is
established, we can use our previous results again. Namely, different
combinations of collisions between an electron and a photon are considered
in Sec. VI\ of \cite{k} with the conclusion that pairs (critical electron,
usual photon) and (critical photon, usual electron) lead to infinitely
growing energies in the centre of mass frame$.$

\section{Conclusion}

Thus, from geometric reasonings, we deduced kinematic properties of
particles moving near the horizon ($N\rightarrow 0$) both for the massive
and massless cases. This revealed the role of "critical" particles as having
special behavior of coefficients of expansion of the four-velocity (or the
wave vector) with respect to the null version of the ZAMO basis. The effect
of infinite grow if the energy in the centre of mass frame depends crucially
on whether $v\rightarrow 1$ or $v\neq 1$ for massive particles near the
horizon and whether $\omega \rightarrow \infty $ or $\omega $ is finite in
the massless case. Once the existence of critical particle is noticed,
further results follow directly from previous works \cite{k}, details of
which are not repeated here.

The present approach revealed the generality of the effect under discussion.
We did not use geodesic equations of motion. Moreover, we even did not use
the existence of Killing vectors and did not assume that the metric
coefficients (\ref{m}) are independent of $t$ and $\phi $. Therefore, the
results have a rather general character. They apply to any surfaces of
infinite redshidt (horizons) which can be characterized by the property $%
N\rightarrow 0$ in metric (\ref{m}). Thus, pure \ geometric approach of \cite%
{cqg} perfectly agrees with the kinematic ones of \cite{k} under very
general circumstances.

This generality means a challenge to attempts to restrict the effect under
discussion in such collisions invoking backreaction or gravitational
radiation \cite{berti}, \cite{ted}. It is not clear, whether and how account
for these factors can restrict the grow of the energy and in what way they
can change the role of critical particles. For better understanding, dynamic
analysis should be combined with geometrical and kinematic approaches.

\end{document}